\documentclass{article}
\usepackage[text={7in,9in},centering]{geometry}
\usepackage{graphicx,epsfig}
\usepackage{cite}
\usepackage{amsmath,amssymb}
\usepackage{hyperref}

\usepackage[usenames]{color}
\usepackage{ulem} 

\definecolor{brown}{rgb}{0.3,0.2,0}

\begin{document}
\title{Electromagnetic radiation detectors based on Josephson junctions: Effective Hamiltonian}
\author{D. V. Anghel, K. Kulikov, Y. M. Galperin, L. S. Kuzmin}
\maketitle

\begin{abstract}
We theoretically analyze two setups of low energy single-photon counters based on Josephson junctions (JJ).
For this, we propose two simple and general models, based on the macroscopic quantum tunneling formalism (MQT).
The first setup is similar to the photon counter based on the ``cold electron bolometer'' (CEB), where the JJ replaces the CEB in the center of the superconducting antenna.
In the second setup, the JJ is capacitively coupled to the antenna.
We derive the Hamiltonians for the two setups and we write the Schr\"odinger equations taking into account both, the antenna and the JJ.
The quantum particles of the MQT models move in two-dimensional potential landscapes, which are parabolic along one direction and may have the form of a washboard potential along another direction.
Such a potential landscape has a series of local minima, separated by saddle points.
If the particle is prepared in the initial state in the metastable ``ground state'' of a local minimum, then the photon absorption causes it to jump into an excited state.
If the excitation energy is bigger than the potential barrier seen by the quantum particle (the difference between the ``ground state'' and the saddle point), the photon is detected.

The models are simple and allow us to do mostly analytical calculations.
We show that the two setups are equivalent form the MQT point of view, since one Hamiltonian can be transformed into the other by changes of variables.
For typical values of the JJ and antenna parameters, the setups may work as counters of photons of wavelengths up to 1~cm, at least.
Dark count rates due to the phase particle tunneling directly from the ground state into the running state have also been evaluated.
\end{abstract}

\section{Introduction} \label{sec_intro}

Photon counters are required for a variety of applications, such as ra\-di\-ation-matter interaction, quantum optics, astrophysics, atomic physics, and quantum information processing~\cite{RevModPhys.78.217.2006.Giazotto, NatPhotonics.3.696.2009.Hadfield, NatPhys.6.663.2010.Johnson, OpticalEngineering.53.081907.2014.Dauler, RevSciInstrum.82.071101.2011.Eisaman, NatPhys.14.546.2018.Kono, PhysRevX.8.021003.2018.Besse, PhysRevLett.120.203602.2018.Royer, Science.361.1239.2018.Opremcak}.
Furthermore, the race for detecting centimeter long wavelength photons was intensified by the proposal that axions -- the elusive particles of the Standard Model that might participate in the formation of Dark Matter~\cite{PhysRevLett.38.1440.1977.Peccei, PhysRevD.16.1791.1977.Peccei, PhysRevLett.40.223.1978.Weinberg, PhysRevLett.40.279.1978.Wilczek} -- could be detected after they decay into such photons when they pass through a region of high magnetic field~\cite{PhysRevLett.51.1415.1983.Sikivie, PhysDarkUniv.15.2212.2017.Barbieri}.
Eventually the most promising technologies that may be employed for detecting photons of centimeter long wavelength are based on superconducting devices~\cite{RevModPhys.78.217.2006.Giazotto, NatPhotonics.3.696.2009.Hadfield, NatPhys.6.663.2010.Johnson, OpticalEngineering.53.081907.2014.Dauler, RevSciInstrum.82.071101.2011.Eisaman, NatPhys.14.546.2018.Kono, PhysRevX.8.021003.2018.Besse, PhysRevLett.120.203602.2018.Royer, Science.361.1239.2018.Opremcak, PhysC.470.2079.2010.Anders, PhysRevLett.107.217401.2011.Chen, PhysRevLett.102.173602.2009.Romero, PhysicaScripta.T137.014004.2009.Romero, PhysicaB.165-166.959.1990.Takayanagi, IntWorkshopSupercondNanoElDev.145.2001.Kuzmin, JPhysConfSer.97.012310.2008.Kuzmin, IEEETransApplSupercond.21.3635.2011.Tarasov, ApplPhysLett.82.293.2003.Anghel, ApplPhysLett.103.142605.2013.Oelsner, PhysRevApplied.7.014012.2017.Oelsner, Nature.431.162.2004.Wallraff, PhysRevA.87.052119.2013.Andersen, PhysRevA.89.033853.2014.Andersen, PhysSolidState.58.2160.2016.Ilichev, PhysRevB.86.174506.2012.Poudel}. However, for detecting axions these detectors have to be coupled to an antenna and should monitor the environment for rare events -- eventually one in a few hours.
For example, the cold electron bolometer (CEB)~\cite{IntWorkshopSupercondNanoElDev.145.2001.Kuzmin, JPhysConfSer.97.012310.2008.Kuzmin, IEEETransApplSupercond.21.3635.2011.Tarasov} has been proposed as a counter~\cite{ApplPhysLett.82.293.2003.Anghel} for photons of wavelengths up to 1~cm~\cite{arXiv181105326.Anghel}.
The CEB is capacitively coupled to an antenna that absorbs the photon.
The energy of the photon is dissipated into the normal metal island of the CEB, causing a rise in the  electrons temperature in the normal metal, which may be detected by the normal metal-insulator-superconductor (NIS) tunnel junctions \cite{ApplPhysLett.65.3123.1994.Nahum, ApplPhysLett.68.1996.1996.Leivo, RevModPhys.78.217.2006.Giazotto}.

Another method to detect microwave photons produced by axions is using Josephson junctions (JJ)~\cite{Phys.Lett.1.251.1962.Josephson, Likharev:book, arXiv0508728.Ingold} as photon counters~\cite{ApplPhysLett.103.142605.2013.Oelsner, PhysRevApplied.7.014012.2017.Oelsner, Nature.431.162.2004.Wallraff, PhysRevA.87.052119.2013.Andersen, PhysRevA.89.033853.2014.Andersen}.
At a bias current lower than the critical current, the JJ may be in the superconducting (non-dissipative) regime, when its dynamics is described along the macroscopic quantum tunneling (MQT) formalism, as a quantum particle in a  washboard potential.
Coupled to an antenna, which can absorb the photon, the quantum particle representing the JJ may be excited and escape over the barrier created by the washboard potential or  may  tunnel through it.
Such a process would put the system into the running state, the JJ becomes dissipative, and the photon is detected.

We shall present two setups for the construction of the JJ detector.
The first setup is based on the design of the photon counter with CEB, as the one proposed in~\cite{ApplPhysLett.82.293.2003.Anghel, arXiv181105326.Anghel}.
In this setup, the CEB is replaced by a current biased JJ, which is placed in the center of a superconducting antenna and is electrically connected in series with it.
In the second setup, the antenna is capacitively connected to the JJ, as shown, for example, in Ref.~\cite{PhysRevA.89.033853.2014.Andersen}.

In this paper, we present the MQT description of whole system in both setups, i.e., we include in the description the JJ, as well as the antenna.
The photon absorption is seen as a single excitation of the system, which may (or may not) drive the quantum particle over the potential barrier.
To simplify the description, so that we can do mostly analytical calculations and capture the essential physical phenomena, we do not describe the environment of the JJ in full detail (i.e., as a transmission line coupled to the JJ~\cite{PhysRevA.89.033853.2014.Andersen}), but we shall model it as a simple LC circuit, coupled to the  JJ.
We observe that the MQT quantum particle moves in a two-dimensional (2D) potential landscape.
Depending on the parameters of the system, the potential energy may look like a 2D washboard potential (which is concave in a direction which is not along the washboard potential), with an infinite number of local minima and saddle points between them.
The absorption of a photon excites the MQT particle, which may be in the metastable ground state of one of the local minima.
If the excitation energy (i.e., the photon's energy) exceeds the difference between the ground state energy (in which the MQT particle stays) and the closest saddle point energy, then the system gets into the running state, the JJ becomes dissipative, and a voltage pulse is detected in the circuit, which counts the photon.
Certainly, the system may get into the running state after the absorption of the photon by tunneling through the potential barrier.
In this paper, we focus on derivation of the effective Hamiltonian of the device. Its quantum dynamics including an interplay between activation and tunneling, as well as thermalization after an absorption event, will be considered elsewhere. Such an analysis will allow estimating the ``dead time" of the detector after counting of a photon.
We shall see that the two experimental setups are equivalent, as one can transform the Hamiltonian of one system into the Hamiltonian of the other by a change of variables.

The paper is organized as follows. In the next section we present the electric circuit diagrams for the two setups and the corresponding quantum descriptions.
By algebraic manipulations, we bring the two Hamiltonian operators to the same form and investigate the properties of the potential energy landscape.
We find analytically the positions of the minima, the saddle points and the difference in energy between consecutive minima and saddle points.
To find the (metastable) ground state energy, we approximate the Hamiltonian around a local minimum with a harmonic potential and calculate the energy of the ground state, to finally calculate the energy required to bring the system into the running state and detect the photon.
We investigate if the system can be realized, using concrete physical parameters.
Last section is reserved for conclusions.

\section{Methods} \label{sec_Form}

The experimental setups that we are going to describe are similar to the  ones presented in Refs.~\cite{ApplPhysLett.82.293.2003.Anghel} (\textit{first setup}) and~\cite{PhysRevA.89.033853.2014.Andersen} (\textit{second setup}).
In the first setup, the CEB is replaced with the JJ which is connected in series with the superconducting antenna.
In the second setup, the JJ is capacitively connected with the antenna.
In both cases, the antenna is described as a simple LC circuit, of inductance $L_A$ and capacitance $C_A$.
Similarly, the JJ in both cases has the parameters $E_c$ (charge energy), capacitance $C_J$, critical current $I_c$, and instantaneous current $I_J$.
The two equivalent circuits are presented in Figs.~\ref{Antena_JJ_detector} and~\ref{Antena_JJ_detector_v2}.

\subsection{First setup} \label{subsec_Conf1}

\begin{figure}[t]
  \centering
  \includegraphics[width=7cm,keepaspectratio=true]{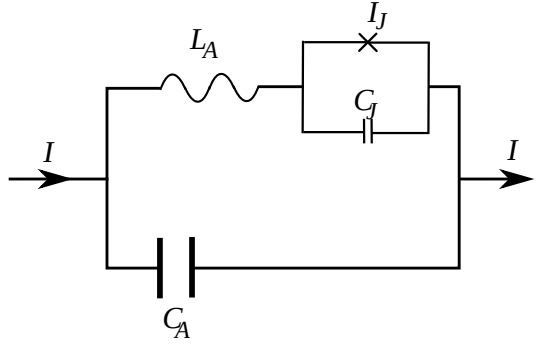}
  \caption{The equivalent circuit of the detector in the \textit{first setup}.
  The current through the JJ is $I_J$ and the capacitance is $C_J$.
  The superconducting antenna  of impedance $L_A$ and capacitance $C_A$ is connected in series with the JJ.}
  \label{Antena_JJ_detector}
\end{figure}

The equivalent electrical circuit of the first setup~\cite{ApplPhysLett.82.293.2003.Anghel} is presented in Fig.~\ref{Antena_JJ_detector}.
The JJ is placed in series with the superconducting antenna and is current biased.
The total current through the circuit is the external current $I$ and the  total voltage is $V_T$.
If the current through the junction, $I_J$, is smaller than the critical current $I_c$, then the JJ is in the superconducting state and we can write $I_J = I_c \sin\phi_J$, where $\phi_J$ is the phase drop on  the junction.
Similarly, the  current through $L_A$ is $I_A$, the charges on the capacitors are $Q_A = V_T C_A$ and $Q_J = V_J C_J$.
Then, the energy of the JJ is~\cite{Likharev:book}
\begin{subequations} \label{EJ_EA_ET}
\begin{equation}
  E_J = \frac{C_J V_J^2}{2} + E_c \left( 1 - \cos\phi_J \right) \equiv \frac{Q_J^2}{2 C_J} + E_c \left( 1 - \cos\phi_J \right)  \label{EJ}
\end{equation}
whereas the energy in the antenna is
\begin{equation}
  E_A = \frac{C_A V_T^2}{2} + \frac{L_A I_A^2}{2} \equiv \frac{Q_A^2}{2 C_A} + \frac{L_A I_A^2}{2}.  \label{EA}
\end{equation}
%
The total energy of the system is
\begin{equation}
  E_T^{(1)} = E_J + E_A .  \label{ET}
\end{equation}
\end{subequations}
Since $E_T^{(1)}$ is not conserved because  of the power supplied from the external circuit, $P_{\text{ext}} \equiv I V_T$, we introduce the overall phase
\begin{equation}
  \phi_T (t) = \frac{2e}{\hbar}  \int_{-\infty}^t dt'\, V_T(t') \ \Leftrightarrow \
  \dot \phi_T = \frac{2e}{\hbar} V_T(t)
  \label{phi_T}
\end{equation}
and write the Hamiltonian
\begin{equation}
  H^{(1)} = E_T^{(1)} - \frac{\hbar}{2e} I \phi_T , \label{def_H}
\end{equation}
which is conserved and may be used in the MQT procedure~\cite{arXiv0508728.Ingold}.

If we denote by $V_A$ the voltage on $L_A$ and since $V_T$ is the voltage on $C_A$, then we have $V_T = V_J + V_A$, which implies
\begin{equation}
  \phi_T = \phi_J + \phi_A  \label{phi_T2}
\end{equation}
where
\begin{equation}
  \phi_A = \frac{2e}{\hbar}  \int_{-\infty}^t dt'\, V_A(t')
  . \label{phi_J}
\end{equation}
Putting together Eqs.~(\ref{EJ_EA_ET})-(\ref{phi_J}), we obtain the total Hamiltonian of the system in terms of $\phi_T$ and $\phi_J$,
\begin{eqnarray}
  H^{(1)} &=&
  \frac{C_A}{2} \left( \frac{\hbar}{2e} \right)^2 \dot\phi_T^2 + \frac{1}{2 L_A} \left( \frac{\hbar}{2e} \right)^2 \phi_T^2 - \frac{\hbar}{2e} I \phi_T \nonumber \\
  && + \frac{ C_J}{2} \left( \frac{\hbar}{2e} \right)^2 \dot\phi_J^2 + E_c \left( 1 - \cos\phi_J \right) + \frac{1}{2 L_A} \left( \frac{\hbar}{2e} \right)^2 \phi_J^2
- \frac{1}{L_A} \left( \frac{\hbar}{2e} \right)^2 \phi_T \phi_J
  . \label{HT2}
\end{eqnarray}
From Eq.~(\ref{HT2}) we calculate the conjugate momenta of the variables $\phi_T$ and $\phi_J$,
\begin{subequations} \label{pA_pJ}
\begin{eqnarray}
  p_T &=& \frac{\partial H^{(1)}}{\partial \dot \phi_T} = C_A \left( \frac{\hbar}{2e} \right)^2 \dot\phi_T = C_A \frac{\hbar}{2e} V_T \equiv \frac{\hbar}{2e} Q_A,  \label{pA} \\
  p_J &=& \frac{\partial H^{(1)}}{\partial \dot \phi_J} = C_J \left( \frac{\hbar}{2e} \right)^2 \dot\phi_J = C_J \frac{\hbar}{2e} V_J \equiv \frac{\hbar}{2e} Q_J .  \label{pJ}
\end{eqnarray}
\end{subequations}
We write the quantum mechanical Hamiltonian as
\begin{eqnarray}
  \hat H^{(1)} &=& - \frac{\hbar^2}{2 C_A} \left( \frac{2e}{\hbar} \right)^2 \frac{\partial^2}{\partial \phi_T^2} + \frac{1}{2 L_A} \left( \frac{\hbar}{2e} \right)^2 \phi_T^2 - \frac{\hbar}{2e} I \phi_T \nonumber \\
  && - \frac{\hbar^2}{2 C_J} \left( \frac{2e}{\hbar} \right)^2 \frac{\partial^2}{\partial\phi_J^2} + E_c \left( 1 - \cos\phi_J \right) + \frac{1}{2 L_A} \left( \frac{\hbar}{2e} \right)^2 \phi_J^2
 - \frac{1}{L_A} \left( \frac{\hbar}{2e} \right)^2 \phi_T \phi_J
  . \label{QM_HT}
\end{eqnarray}

It is more convenient to work with dimensionless quantities.
If $(L_AC_A)^{-1/2}$ is the resonant frequency of the antenna, which should be the same as the frequency of the absorbed photon, we may re-scale the total Hamiltonian by what we shall call \textit{the photon's energy} $\hbar/\sqrt{L_AC_A}$ and introduce the dimensionless Hamiltonian
\begin{eqnarray}
  \hat H_d^{(1)} &=& \frac{\sqrt{L_AC_A}}{\hbar} \hat H^{(1)} = - \frac{1}{2} \frac{(2e)^2}{\hbar} \sqrt{\frac{L_A}{C_A}} \frac{\partial^2}{\partial \phi_T^2}
  + \frac{1}{2} \frac{\hbar}{(2e)^2} \sqrt{\frac{C_A}{L_A}} \phi_T^2
  - \frac{\sqrt{L_AC_A}}{2e} I \phi_T \nonumber \\
  && - \frac{1}{2} \frac{(2e)^2}{\hbar} \frac{\sqrt{L_AC_A}}{C_J} \frac{\partial^2}{\partial\phi_J^2}
  + \frac{I_c \sqrt{L_AC_A}}{2e} \left( 1 - \cos\phi_J \right)
  + \frac{1}{2} \frac{\hbar}{(2e)^2} \sqrt{\frac{C_A}{L_A}} \phi_J^2
   - \frac{\hbar}{(2e)^2} \sqrt{\frac{C_A}{L_A}} \phi_T \phi_J . \nonumber
\end{eqnarray}
We denote
$$\tilde\phi_T \equiv  \left(\frac{\hbar}{(2e)^2} \sqrt{\frac{C_A}{L_A}}\right)^{1/2} \!\phi_T \ \text{and} \  \phi_J' \equiv \left(\frac{\hbar}{(2e)^2} \sqrt{\frac{C_J}{L_A}}\right)^{1/2}\! \phi_J $$
to write $\hat H_d^{(1)}$ in the simpler form
\begin{eqnarray}
  \hat H_d^{(1)} &=& - \frac{1}{2} \frac{\partial^2}{\partial \tilde\phi_T^2}
  + \frac{1}{2} \tilde\phi_T^2
  - \left( \frac{L_A^3 C_A}{\hbar^2} \right)^{1/4} I \tilde\phi_T \nonumber \\
  && + \sqrt{\frac{C_A}{C_J}} \left\{ - \frac{1}{2} \frac{\partial^2}{\partial(\phi_J')^2}
  + \frac{I_c \sqrt{L_AC_J}}{2e} \left[ 1 - \cos\left( \frac{\phi_J'}{\sqrt{\frac{\hbar}{(2e)^2} \sqrt{\frac{C_J}{L_A}}}} \right) \right]
  + \frac{1}{2} (\phi_J')^2 \right\}
   - \left( \frac{C_A}{C_J} \right)^{1/4} \tilde\phi_T \phi_J'
  . \label{QM_HT_v2}
\end{eqnarray}
Introducing the variable $\tilde\phi_J \equiv (C_J/C_A)^{1/4} \phi_J'$ and the notations $\omega_A \equiv (L_AC_A)^{-1/2}$, $Z_0 \equiv \frac{\hbar}{(2e)^2}$, $Z_A \equiv \sqrt{L_A/C_A}$, we write the Hamiltonian~(\ref{QM_HT_v2}) as
\begin{eqnarray}
  \hat H_d^{(1)} &=& - \frac{1}{2} \frac{\partial^2}{\partial \tilde\phi_T^2}
  - \frac{1}{2} \frac{\partial^2}{\partial \tilde\phi_J^2}
  + \frac{1}{2} \left( \tilde\phi_T - \sqrt{\frac{C_A}{C_J}} \tilde\phi_J \right)^2
  - \left( \frac{L_A^3 C_A}{\hbar^2} \right)^{1/4} I \tilde\phi_T
   + \frac{I_c \sqrt{L_AC_A}}{2e} \left[ 1 - \cos\left( \frac{\tilde\phi_J}{\sqrt{\frac{\hbar}{(2e)^2} \frac{C_J}{\sqrt{L_AC_A}}}} \right) \right] \nonumber \\
  %
  %
  &=& - \frac{1}{2} \frac{\partial^2}{\partial \tilde\phi_T^2}
  - \frac{1}{2} \frac{\partial^2}{\partial \tilde\phi_J^2}
  + \frac{1}{2} \left(\tilde\phi_T - \sqrt{\frac{C_A}{C_J}} \tilde\phi_J - \frac{1}{2e \omega_A} \sqrt{\frac{Z_A}{Z_0}} I \right)^2 \nonumber \\
  && \quad + \frac{I_c}{2e \omega_A} \left[ 1 - \cos\left( \frac{\tilde\phi_J}{\sqrt{Z_0 C_J \omega_A}} \right) \right]
  - \frac{1}{2e \omega_A} \frac{1}{\sqrt{Z_0 C_J \omega_A}} I \tilde\phi_J
  - \frac{1}{2} \frac{1}{(2e)^2 \omega_A^2} \frac{Z_A}{Z_0} I^2.  \label{QM_HT_v5}
\end{eqnarray}

We observe that the quantum particle moves in a 2D potential, which is parabolic along the direction $\tilde\phi_T - \sqrt{C_A/C_J} \tilde\phi_J$ and has a washboard shape along the direction $\tilde\phi_J$.

\subsection{Second setup} \label{subsec_Conf2}

\begin{figure}[t]
  \centering
  \includegraphics[width=7cm,keepaspectratio=true]{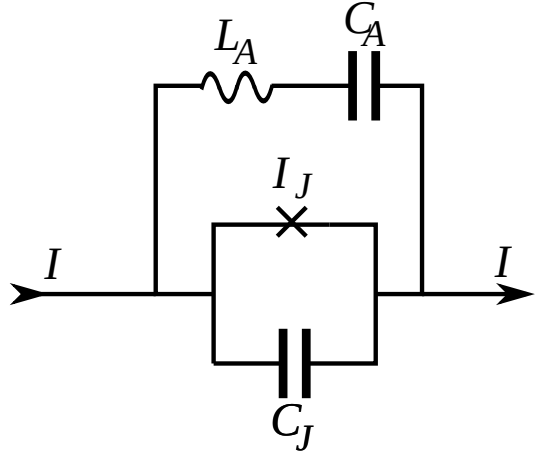}
  \caption{The equivalent circuit of the detector in the \textit{second setup}.
  The current through the JJ is $I_J$ and the capacitance is $C_J$.
  The superconducting antenna, of impedance $L_A$ and capacitance $C_A$ is capacitively connected with the JJ.
  }
  \label{Antena_JJ_detector_v2}
\end{figure}

We analyze now the second setup, similar to the one in Ref.~\cite{PhysRevA.89.033853.2014.Andersen}.
A simplified version of the circuit is depicted in Fig.~\ref{Antena_JJ_detector_v2}.
In this case, the Hamiltonian for the JJ remains the same as Eq.~(\ref{EJ}), with the modification that now the voltage on the JJ capacitor is the total voltage, i.e., $V_J \equiv V_T$.
On the other hand, the energy of the antenna is
\begin{equation}
  E_A = \frac{C_A V_A^2}{2} + \frac{L_A I_A^2}{2} \equiv \frac{Q_A^2}{2 C_A} + \frac{L_A I_A^2}{2} . \label{EA_v2}
\end{equation}
The total energy of the  system is given again by Eq.~(\ref{ET}).
The total phase and the Hamiltonian are~(\ref{phi_T}) and~(\ref{def_H}), respectively.
If, like in Section~\ref{subsec_Conf1}, $I_A$ is the current through the antenna, $I_J$ is the current through the JJ and $I_{C_J}$ is the current through the capacitance of the JJ, then the current conservation reads
\begin{equation}
  I = I_A + I_J + I_{C_J} . \label{cons_I}
\end{equation}
Equation (\ref{cons_I}) couples the Hamiltonian of the antenna with the Hamiltonian of the JJ.

Denoting by $V_{L_A}$ the voltage on the antenna inductance and by $V_{C_A}$ the voltage on the antenna capacitance, we define the phases
\begin{subequations} \label{def_phiLA_phiCA_phiCJ}
\begin{eqnarray}
  \phi_{L_A} &\equiv& \frac{2e}{\hbar} \int_{-\infty}^t dt' V_{L_A} (t') = \frac{2e}{\hbar} \int_{-\infty}^t dt' \left[ V_T (t') - \frac{Q_{C_A}(t')}{C_A} \right] , \label{def_phiLA} \\
  \phi_{C_A} &\equiv& \frac{2e}{\hbar} \int_{-\infty}^t dt' V_{C_A} (t') = \frac{2e}{\hbar} \int_{-\infty}^t dt' \frac{Q_{C_A}(t')}{C_A} , \label{def_phiCA} \\
  \phi_J &=& \frac{2e}{\hbar} \int_{-\infty}^t dt' V_J (t') \equiv \frac{2e}{\hbar} \int_{-\infty}^t dt' V_T (t') \equiv \phi_T , \label{def_phiCJ}
\end{eqnarray}
\end{subequations}
which have to be embedded into the total Hamiltonian of the system,
\begin{equation}
  E_T^{(2)} = \frac{Q_T^2}{2C_J} + E_c (1 - \cos\phi_J) + \frac{Q_{C_A}^2}{2 C_A} + \left( \frac{\hbar}{2e} \right)^2 \frac{\phi_{L_A}^2}{2L_A} - \frac{\hbar}{2e} I \phi_T , \label{EA_v2_2}
\end{equation}
where $Q_T$ is the charge on the JJ capacitance.
Using the relation $\phi_{L_A} \equiv \phi_T - \phi_{C_A}$, from~(\ref{EA_v2_2}) we obtain
\begin{eqnarray}
  E_T^{(2)} &=& \frac{Q_T^2}{2C_J} + E_c (1 - \cos\phi_T) + \left( \frac{\hbar}{2e} \right)^2 \frac{\phi_T^2}{2L_A} - \frac{\hbar}{2e} I \phi_T
   + \frac{Q_{C_A}^2}{2 C_A} + \left( \frac{\hbar}{2e} \right)^2 \frac{\phi_{C_A}^2}{2L_A} - \left( \frac{\hbar}{2e} \right)^2 \frac{\phi_T \phi_{C_A}}{L_A}.
  \label{EA_v2_3}
\end{eqnarray}
From Eq.~(\ref{EA_v2_3}) we obtain the Hamiltonian
\begin{eqnarray}
  \hat H^{(2)} &=& - \frac{(2e)^2}{2C_J} \frac{\partial^2}{\partial \phi_T^2} + E_c (1 - \cos\phi_T) + \left( \frac{\hbar}{2e} \right)^2 \frac{\phi_T^2}{2L_A} - \frac{\hbar}{2e} I \phi_T
  \nonumber \\ && \quad
   - \frac{(2e)^2}{2 C_A} \frac{\partial^2}{\partial \phi_{C_A}^2} + \left( \frac{\hbar}{2e} \right)^2 \frac{\phi_{C_A}^2}{2L_A} - \left( \frac{\hbar}{2e} \right)^2 \frac{\phi_T \phi_{C_A}}{L_A} .
  \label{H2}
\end{eqnarray}
To make the Hamiltonian dimensionless, we divide Eq.~(\ref{H2}) by the photon's energy and define
\begin{eqnarray}
  \hat H_d^{(2)} &\equiv& 
  \frac{\hat H^{(2)}}{\hbar \omega_A}  = - \frac{1}{2} \frac{1}{Z_0 C_J \omega_A} \frac{\partial^2}{\partial \phi_T^2}
  + \frac{I_c}{2e \omega_A} (1 - \cos\phi_T) \nonumber \\
  %
  %
  && + \frac{1}{2} \frac{Z_0}{Z_A} \phi_T^2
  - \frac{1}{2e \omega_A} I \phi_T - \frac{1}{2} \frac{Z_A}{Z_0} \frac{\partial^2}{\partial \phi_{C_A}^2}
   + \frac{1}{2} \frac{Z_0}{Z_A} \phi_{C_A}^2
  - \frac{Z_0}{Z_A} \phi_T \phi_{C_A}.
  \label{H2_ad}
\end{eqnarray}
Introducing the variables $\tilde \phi_T = \sqrt{Z_0 C_J \omega_A} \phi_T$
and $\tilde\phi_{C_A} = \sqrt{\frac{Z_0}{Z_A}} \phi_{C_A}$ we obtain
\begin{eqnarray}
  \hat H_d^{(2)} &=& - \frac{1}{2} \frac{\partial^2}{\partial \tilde\phi_T^2}
  - \frac{1}{2} \frac{\partial^2}{\partial \tilde\phi_{C_A}^2}
   + \frac{1}{2} \left( \tilde\phi_{C_A} - \sqrt{\frac{C_A}{C_J}} \tilde\phi_T \right)^2
 \nonumber \\
 %
 %
 && + \frac{I_c}{2e \omega_A} \left[1 - \cos\left( \frac{\tilde \phi_T}{\sqrt{Z_0 C_J\omega_A}} \right) \right] - \frac{1}{2e \omega_A} \frac{1}{\sqrt{Z_0 C_J \omega_A}} I \tilde\phi_T. \label{H2_ad_v2}
\end{eqnarray}
We observe that (making the identification $\tilde\phi_{C_A} \equiv \tilde\phi_J$) the Hamiltonian~(\ref{H2_ad_v2}) is equivalent to the Hamiltonian~(\ref{QM_HT_v5}), except that in~(\ref{QM_HT_v5}) the potential energy is shifted by the constant value $-\left( L_A^3 C_A/2\hbar^2 \right)^{1/2} I^2$ and is translated in the direction $\tilde\phi_J - \sqrt{C_A/C_J} \tilde\phi_T$ by $- \sqrt{Z_A/Z_0} I / (2e \omega_A)$. 
\label{page_ref_dif_U}
These differences are irrelevant for the dynamics of the quantum particle and we shall see that all the physical results of the two setups are identical.


\section{Results} \label{sec_results}

Having derived the Hamiltonians for the two configurations, we can calculate the local minima, the saddle points and the energy required to excite the system over the saddle point.

\subsection{Local minima and saddle points for the first setup} \label{subsec_min_set1}

The particle described by the Hamiltonian $\hat H_d^{(1)}$~(\ref{QM_HT_v5}) of the first setup is in a potential energy
\begin{eqnarray}
  U^{(1)}_d &\equiv& \frac{1}{2} \left[\tilde\phi_T - \sqrt{\frac{C_A}{C_J}} \tilde\phi_J - \frac{1}{2e \omega_A} \sqrt{\frac{Z_A}{Z_0}} I \right]^2
  - \frac{1}{2} \frac{1}{(2e)^2 \omega_A^2} \frac{Z_A}{Z_0} I^2 \label{QM_U1} \\
  && + \frac{I_c}{2e \omega_A} \left[ 1 - \cos\left( \frac{\tilde\phi_J}{\sqrt{Z_0 C_J\omega_A}} \right) \right]
  - \frac{1}{2e \omega_A} \frac{1}{\sqrt{Z_0 C_J \omega_A}} I \tilde\phi_J
  . \nonumber
\end{eqnarray}
As noticed before, along the $\tilde\phi_J$ direction we have a washboard potential, whereas along the $\tilde\phi_T - \sqrt{C_A/C_J} \tilde\phi_J$ direction we have a harmonic potential.
The minima and the saddle points of these potentials are found by calculating the derivatives
\begin{subequations} \label{ders_U1}
\begin{eqnarray}
  \frac{\partial U^{(1)}_d}{\partial \tilde\phi_T} &=& \left( \tilde\phi_T - \sqrt{\frac{C_A}{C_J}} \tilde\phi_J \right)
  - \frac{1}{2e \omega_A} \sqrt{\frac{Z_A}{Z_0}} I , \label{dU1_dft} \\
  %
  \frac{\partial^2 U^{(1)}_d}{\partial (\tilde\phi_T)^2} &=& 1 , \label{dU12_dft2} \\
  \frac{\partial^2 U^{(1)}_d}{\partial \tilde\phi_T \partial \tilde\phi_J} &=& - \sqrt{\frac{C_A}{C_J}} , \label{dU12_dft_dfJ} \\
  %
  \frac{\partial  U^{(1)}_d}{\partial \tilde\phi_J} &=& - \sqrt{\frac{C_A}{C_J}} \left( \tilde\phi_T - \sqrt{\frac{C_A}{C_J}} \tilde\phi_J \right)
   + I_c \frac{1}{2e \omega_A} \frac{1}{\sqrt{Z_0 C_J \omega_A}} \sin\left( \frac{\tilde\phi_J}{\sqrt{Z_0 C_J \omega_A}} \right) , \label{dU1_dfJ} \\
  %
  \frac{\partial^2  U^{(1)}_d}{\partial \tilde\phi_J^2} &=& \frac{C_A}{C_J}
  + \frac{I_c}{2e Z_0 C_J \omega_A^2} \cos\left( \frac{\tilde\phi_J}{\sqrt{Z_0 C_J \omega_A}} \right) . \label{dU12_dfJ2}
\end{eqnarray}
\end{subequations}
From Eqs.~(\ref{dU1_dft}) and (\ref{dU1_dfJ}) we obtain the relations which have to be satisfied in the local minima and in the saddle points,
\begin{subequations} \label{min_saddle}
\begin{eqnarray}
  \frac{1}{2e \omega_A} \sqrt{\frac{Z_A}{Z_0}} I &=& \left( \tilde\phi_T - \sqrt{\frac{C_A}{C_J}} \tilde\phi_J \right) \label{min_saddle_ft} \\
  %
  {\rm and} \qquad \frac{I}{I_c} &=& \sin\left( \frac{\tilde\phi_J}{\sqrt{Z_0 C_J \omega_A}} \right) . \label{min_saddle_fj}
\end{eqnarray}
\end{subequations}
As expected, we observe that Eq.~(\ref{min_saddle_fj}) has a solution if and only if $-1 < I/I_c < 1$.
If we choose only positive currents ($I\ge 0$), then the local minima are at
\begin{subequations} \label{min_C1}
\begin{eqnarray}
  \tilde\phi_{J,min} &=& 
  \sqrt{Z_0 C_J \omega_A}
  \left[ \arcsin\left(\frac{I}{I_c}\right) + 2 n\pi \right] \quad {\rm and}  \label{min_C1_fj} \\
  \tilde\phi_{T, min} &=& \frac{1}{2e \omega_A} \sqrt{\frac{Z_A}{Z_0}} I +
  \sqrt{\frac{Z_0}{Z_A}} \left[ \arcsin\left(\frac{I}{I_c}\right) + 2n\pi \right] \label{min_C1_ft}
\end{eqnarray}
\end{subequations}
(where $n$ is an integer), whereas the saddle points are at
\begin{subequations} \label{sad_C1}
\begin{eqnarray}
  \tilde\phi_{J,saddle} &=& \sqrt{Z_0 C_J \omega_A} \left[(2n+1)\pi - \arcsin\left(\frac{I}{I_c}\right) \right] \quad {\rm and}  \label{sad_C1_fj} \\
  \tilde\phi_{T, saddle} &=& \frac{I}{2e \omega_A} \sqrt{\frac{Z_A}{Z_0}}
  + \sqrt{\frac{Z_0}{Z_A}} \left[ (2n+1)\pi - \arcsin\left(\frac{I}{I_c}\right) \right] . \label{sad_C1_ft}
\end{eqnarray}
\end{subequations}
The second derivatives in the minima are
\begin{subequations} \label{ders_U1_min}
\begin{eqnarray}
  \frac{\partial^2 U^{(1)}_d}{\partial (\tilde\phi_T)^2} &=& 1 , \label{dU12_dft2_min} \\
  \frac{\partial^2 U^{(1)}_d}{\partial \tilde\phi_T \partial \tilde\phi_J} &=& - \sqrt{\frac{C_A}{C_J}} = \frac{\partial^2 U^{(1)}_d}{\partial \tilde\phi_J \partial \tilde\phi_T} , \label{dU12_dft_dfJ_min} \\
  \frac{\partial^2  U^{(1)}_d}{\partial \tilde\phi_J^2} &=& \frac{C_A}{C_J}
  + \frac{I_c}{2e Z_0 C_J \omega_A^2} \sqrt{1 - \frac{I^2}{I_c^2}} . \label{dU12_dfJ2_min}
\end{eqnarray}
\end{subequations}
The values of the potential energy at the minima and the saddle points are
\begin{subequations} \label{QM_U1_ms}
\begin{eqnarray}
  U^{(1)}_{d,min}(n) &=& - \frac{1}{2} \frac{1}{(2e)^2 \omega_A^2} \frac{Z_A}{Z_0} I^2 + \frac{I_c}{2e \omega_A} \left[ 1 - \sqrt{1 - \frac{I^2}{I_c^2}} \right]
   - \frac{I}{2e \omega_A} \left[ \arcsin\left(\frac{I}{I_c}\right) + 2 n\pi \right], \\  \label{QM_U1_min}
  U^{(1)}_{d,saddle}(n) &=& - \frac{1}{2} \frac{1}{(2e)^2 \omega_A^2} \frac{Z_A}{Z_0} I^2
  + \frac{I_c}{2e \omega_A} \left[ 1 + \sqrt{1 - \frac{I^2}{I_c^2}} \right]
   - \frac{I}{2e \omega_A} \left[ (2n+1)\pi -  \arcsin\left(\frac{I}{I_c}\right) \right] , \label{QM_U1_saddle}
\end{eqnarray}
\end{subequations}
respectively.
The energy difference between a saddle point and the closest local minimum is
\begin{equation}
  U^{(1)}_{d,saddle}(n) - U^{(1)}_{d,min}(n) = \frac{I_c}{e \omega_A} \left\{ \sqrt{1 - \frac{I^2}{I_c^2}} - \frac{I}{I_c} \left[ \frac{\pi}{2} -  \arcsin\left(\frac{I}{I_c}\right) \right] \right\}. \label{QM_U1_diff}
\end{equation}

\subsection{Local minima and saddle points for the second setup} \label{subsec_min_set2}

For the second configuration, from Eq.~(\ref{H2_ad_v2}) we obtain the potential energy
\begin{eqnarray}
  U_d^{(2)} &=& \frac{1}{2} \left( \tilde\phi_{C_A} - \sqrt{\frac{C_A}{C_J}} \tilde\phi_T \right)^2
  + \frac{I_c}{2e \omega_A} \left[1 - \cos\left( \frac{\tilde \phi_T}{\sqrt{Z_0 C_J \omega_A}} \right) \right] - \frac{I}{2e \omega_A} \frac{1}{\sqrt{Z_0 C_J \omega_A}} \tilde\phi_T .  \label{QM_U2}
\end{eqnarray}
Since $U_d^{(2)}$ has the same shape as $U_d^{(1)}$, but in different variables, all the analysis is similar to the one in Section~\ref{subsec_min_set1}.
For this reason, we state here only the main results.
%
%
Along the $\tilde\phi_T$ direction we have a washboard potential, which has local minima if and only if $-1 < I/I_c < 1$. We work again with $I \ge 0$ and we have the coordinates
\begin{subequations} \label{U2_min_C1}
\begin{eqnarray}
  \tilde\phi_{T,min} &=& \sqrt{Z_0 C_J \omega_A} \left[ \arcsin\left(\frac{I}{I_c}\right) + 2n\pi \right] \quad {\rm and}  \label{U2_min_C1_fj} \\
  \tilde\phi_{C_A, min} &=& \sqrt{\frac{Z_0}{Z_A}} \left[ \arcsin\left(\frac{I}{I_c}\right) + 2n\pi \right]  \label{U2_min_C1_ft}
\end{eqnarray}
\end{subequations}
for local minima, and
\begin{subequations} \label{U2_sad_C1}
\begin{eqnarray}
  \tilde\phi_{T,saddle} &=& \sqrt{Z_0 C_J \omega_A} \left[(2n+1)\pi - \arcsin\left(\frac{I}{I_c}\right) \right] \quad {\rm and}  \label{U2_sad_C1_fj} \\
  \tilde\phi_{C_A, saddle} &=& \sqrt{\frac{Z_0}{Z_A}} \left[ (2n+1)\pi - \arcsin\left(\frac{I}{I_c}\right) \right]  \label{U2_sad_C1_ft}
\end{eqnarray}
\end{subequations}
for saddle points.
The second derivatives in the minimum energy points are identical with the ones calculated in Section~\ref{subsec_min_set1} (but in different coordinates), namely
\begin{subequations} \label{ders_U2_min}
\begin{eqnarray}
  \frac{\partial^2 U^{(2)}_d}{\partial \tilde\phi_{C_A}^2} &=& 1 , \label{dU22_dft2_min} \\
  \frac{\partial^2 U^{(2)}_d}{\partial \tilde\phi_{C_A} \partial \tilde\phi_T} &=& - \sqrt{\frac{C_A}{C_J}} = \frac{\partial^2 U^{(2)}_d}{\partial \tilde\phi_T \partial \tilde\phi_{C_A}} , \label{dU22_dft_dfJ_min} \\
  \frac{\partial^2  U^{(2)}_d}{\partial \tilde\phi_T^2} &=& \frac{C_A}{C_J}
  + \frac{I_c}{2e Z_0 C_J \omega_A^2} \sqrt{1 - \frac{I^2}{I_c^2}} .  \label{dU22_dfJ2_min}
\end{eqnarray}
\end{subequations}
The values of the energy at the minima and at the saddle points are
\begin{subequations} \label{QM_U2_ms}
\begin{equation}
  U^{(2)}_{d,min}(n) = \frac{I_c}{2e \omega_A} \left[ 1 - \sqrt{1 - \frac{I^2}{I_c^2}} \right] - \frac{I}{2e \omega_A} \left[ \arcsin\left(\frac{I}{I_c}\right) + 2 n\pi \right] \label{QM_U2_min}
\end{equation}
and
\begin{eqnarray}
  U^{(2)}_{d,saddle}(n) &=& \frac{I_c}{2e \omega_A} \left[ 1 + \sqrt{1 - \frac{I^2}{I_c^2}} \right]
   - \frac{I}{2e \omega_A} \left[ (2n+1)\pi -  \arcsin\left(\frac{I}{I_c}\right) \right] , \label{QM_U2_saddle}
\end{eqnarray}
\end{subequations}
respectively.
As expected,
$U^{(2)}_{d,min}(n) - U^{(1)}_{d,min}(n) = U^{(2)}_{d,saddle}(n) - U^{(1)}_{d,saddle}(n) = (Z_A/Z_0) I^2 / [2(2e\omega_A)^2]$, which implies that $U^{(2)}_{d,saddle}(n) - U^{(2)}_{d,min}(n) = U^{(1)}_{d,saddle}(n) - U^{(1)}_{d,min}(n)$ (see Eq.~\ref{QM_U1_diff}).

\section{Discussion} \label{sec_discussions}

\begin{figure}[t]
  \centering
  \includegraphics[width=6cm]{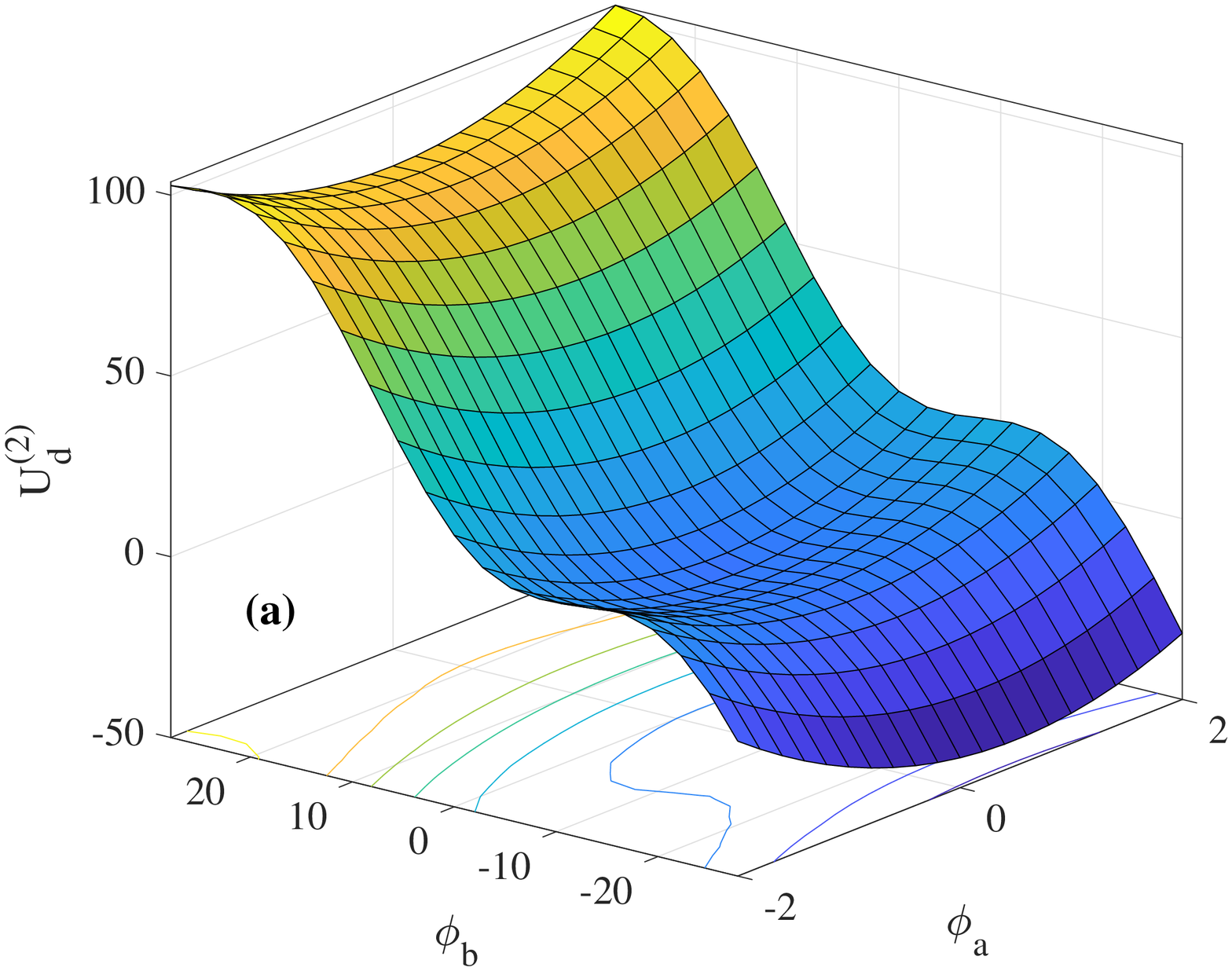}
  \includegraphics[width=6cm]{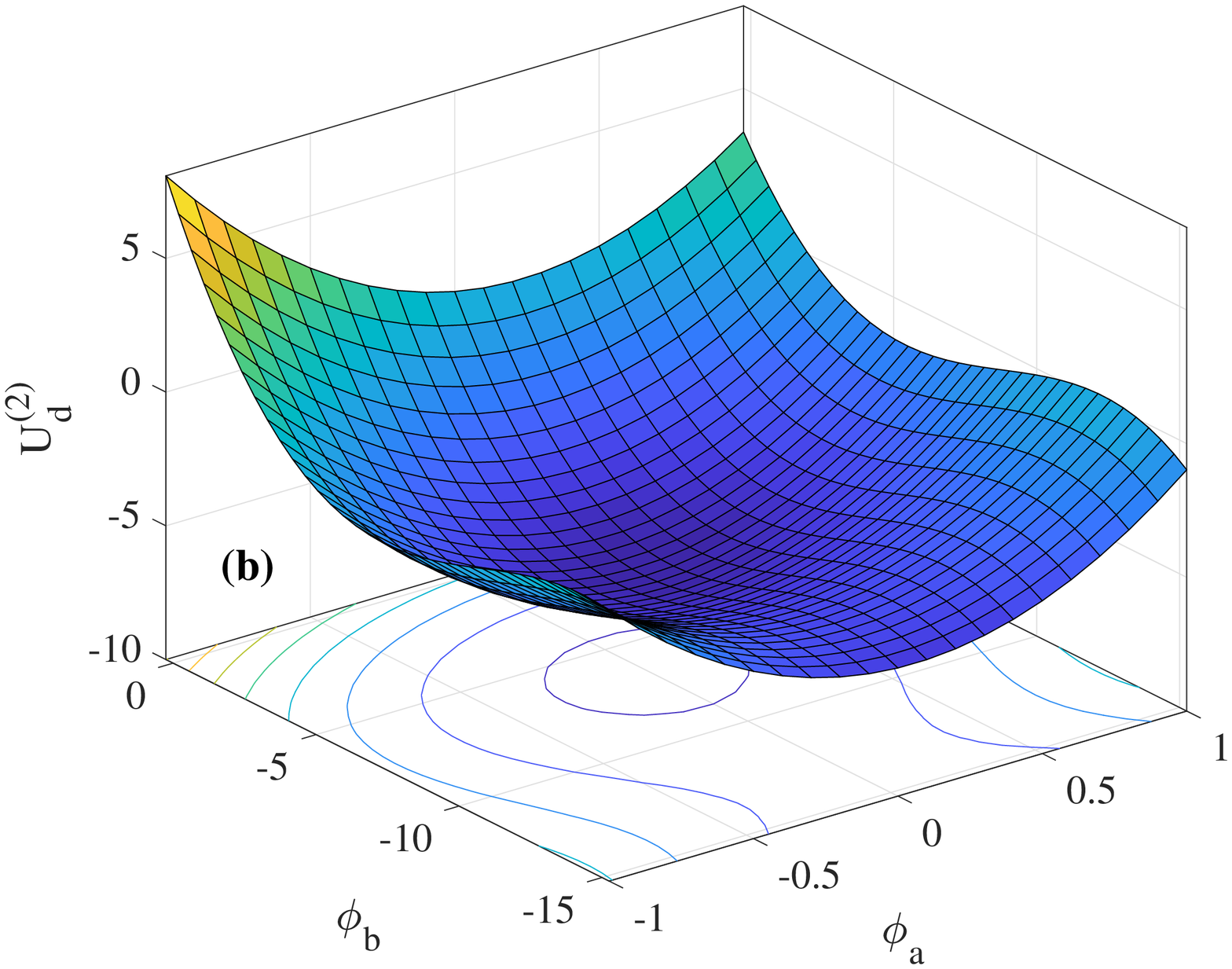}
  \caption{The potential energy $U_d^{(2)}$ in the rotated system of coordinates $(\phi_1, \phi_2)$. The parameters take typical values $I_c = 1~{\rm \mu A}$, $C_A = C_J = 1$~pF~\cite{ApplPhysLett.103.142605.2013.Oelsner, PhysRevApplied.7.014012.2017.Oelsner}, and $\omega_A/(2\pi) = 30$~GHz. The value $I/I_c \approx 0.85$ will be justified later.
  One can observe that the potential is very asymmetric in the variables $\phi_1$ and $\phi_2$.
  The plot in (b) is a detail of the plot in (a), emphasizing a local minimum.}
  \label{U_2d}
\end{figure}

Although the Hamiltonians~(\ref{QM_HT}) and (\ref{H2}), that describe the two schematic setups of Figs.~\ref{Antena_JJ_detector} and~\ref{Antena_JJ_detector_v2}, respectively, are apparently different, we saw in Section~\ref{sec_results} that, by dividing them by $\hbar/\sqrt{L_AC_A}$ and changing the variables, they can be reduced to dimensionless Hamiltonians that represent quantum particles of unit mass, in potential landscapes which are translated with respect to each-other.
A small part of $U_d^{(2)}$ is presented in Fig.~\ref{U_2d}. The system of coordinates is rotated to $(\phi_a, \phi_b)$, defined as
\begin{equation}
  \phi_a \equiv \frac{\tilde\phi_{C_A} - \sqrt{C_A/C_J} \tilde\phi_T}{\sqrt{1 + C_A/C_J}} \quad {\rm and} \quad
  \phi_b \equiv \frac{\sqrt{C_A/C_J} \tilde\phi_{C_A} + \tilde\phi_T}{\sqrt{1 + C_A/C_J}} ,
  \label{vars_f1_f2_H2}
\end{equation}
because in the original coordinates the local potential well is too narrow to be clearly seen.

\begin{figure}[t]
  \centering
  \includegraphics[width=6cm,keepaspectratio=true]{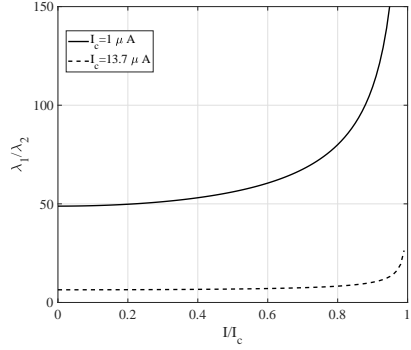}
  \caption{The ratio $\lambda_1/\lambda_2$, for $I_c = 1~{\rm \mu A}$ (solid line), $I_c = 13.7~{\rm \mu A}$ (dashed line), $C_A = C_J = 1$~pF, and $\omega_A / (2\pi) = 30~{\rm GHz}$ (notice that $\lambda_2 \to 0$ when $I/I_c \nearrow 1$).}
  \label{lambda1_lambda2}
\end{figure}

In the initial state (before the absorption of the photon), the quantum particle may sit in the metastable ground state of the local minimum represented in Fig.~\ref{U_2d}~(b).
The energy difference between the local minimum and the nearest saddle point is the same in both setups and can be obtained from Eq.~(\ref{QM_U1_diff}), which gives
\begin{eqnarray}
  \epsilon_{ph}^{(1)} &\equiv& \hbar \omega_A ( U^{(2)}_{d,saddle} - U^{(2)}_{d,min} ) = \hbar \omega_A ( U^{(1)}_{d,saddle} - U^{(1)}_{d,min} )
  \nonumber \\ &=&
   4e I_c Z_0 
   \left\{ \sqrt{1 - \frac{I^2}{I_c^2}} - \frac{I}{I_c} \left[ \frac{\pi}{2} -  \arcsin\left(\frac{I}{I_c}\right) \right] \right\} .
  \label{eps_ph_1}
\end{eqnarray}
To calculate the energy of the photon required to excite the system from the metastable ground state of a local minimum over the saddle point, we approximate the potential energy in the local minimum with a harmonic potential.
The second derivatives of the potential energies $U^{(1)}_d$ and $U^{(2)}_d$ in the local minima are identical, Eqs.~(\ref{ders_U1_min}) and (\ref{ders_U2_min}), so we can analyze only one of them. 
If a local minimum is located at $(\tilde\phi_{J, min}, \tilde\phi_{T, min})$, we denote $\phi_1 \equiv \tilde\phi_{J} - \tilde\phi_{J, min}$ and $\phi_2 \equiv \tilde\phi_{T} - \tilde\phi_{T, min}$.
In the new variables and in the harmonic approximation around $(\tilde\phi_{J, min}, \tilde\phi_{T, min})$, both Hamiltonians $H^{(1)}_d$ and $H^{(2)}_d$ become (neglecting constant terms)
\begin{eqnarray}
  \hat H_d &\equiv& - \frac{1}{2} \frac{\partial^2}{\partial \phi_1^2}
  - \frac{1}{2} \frac{\partial^2}{\partial \phi_2^2}
  + \frac{1}{2} \left( \frac{C_A}{C_J} + \frac{I_c}{2e Z_0C_J \omega_A^2} \sqrt{1 - \frac{I^2}{I_c^2}} \right) \phi_1^2
   - \sqrt{\frac{C_A}{C_J}} \phi_1 \phi_2 + \frac{1}{2} \phi_2^2 . \label{H1_harm}
\end{eqnarray}
Denoting
\begin{subequations} \label{defs_aD}
\begin{eqnarray}
  a &\equiv& \frac{C_A}{C_J} + \frac{I_c}{2e Z_0C_J \omega_A^2} \sqrt{1 - \frac{I^2}{I_c^2}}
  \quad {\rm and} \label{def_a} \\
  D &\equiv& \sqrt{(a-1)^2 + 4 \frac{C_A}{C_J}} \label{def_D}
\end{eqnarray}
\end{subequations}
we define the orthogonal vectors
\begin{equation}
  v_1 \equiv \frac{\phi_1 + \sqrt{\frac{C_J}{C_A}} \frac{1-a+D}{2} \phi_2}{\sqrt{1 + \frac{C_J}{C_A} \frac{(1-a+D)^2}{4}}} \quad {\rm and} \quad
  v_2 \equiv \frac{\phi_1 + \sqrt{\frac{C_J}{C_A}} \frac{1-a-D}{2} \phi_2}{\sqrt{1 + \frac{C_J}{C_A} \frac{(1-a-D)^2}{4}}},
  \label{defs_v1v2}
\end{equation}
which diagonalize the Hamiltonian~(\ref{H1_harm}), bringing it to the form
\begin{subequations} \label{H_v1v2}
\begin{eqnarray}
  \hat H_d &=& - \frac{1}{2} \frac{\partial^2}{\partial v_1^2}
  - \frac{1}{2} \frac{\partial^2}{\partial v_2^2}
  + \frac{1}{2} \lambda_1 v_1^2 + \frac{1}{2} \lambda_2 v_2^2  , \label{H1_harm_diag}
\end{eqnarray}
where
\begin{equation}
  \lambda_1 \equiv \frac{1}{2} (1 + a + D) \quad {\rm and} \quad
  \lambda_2 \equiv \frac{1}{2} (1 + a - D) . \label{lambdas}
\end{equation}
The Hamiltonian $\hat H_d$ of Eq.~(\ref{H1_harm_diag}) may be split into two Hamiltonians, $\hat H_d  \equiv \hat H_{dA} + \hat H_{dB}$, where
\begin{equation}
  \hat H_{dA} \equiv - \frac{1}{2} \frac{\partial^2}{\partial v_1^2}
  + \frac{1}{2} \lambda_1 v_1^2 \quad {\rm and} \quad
  \hat H_{dB} \equiv - \frac{1}{2} \frac{\partial^2}{\partial v_2^2}
  + \frac{1}{2} \lambda_2 v_2^2 .
  \label{HAB_harm}
\end{equation}
\end{subequations}
We observe that at $I/I_c = 1$, $\lambda_1 = 2$ and $\lambda_2 = 0$ for any values of the parameters.
For $I/I_c \in [0,1)$, the ratio $\lambda_1/ \lambda_2$ is plotted in Fig.~\ref{lambda1_lambda2}, for two values of $I_c$.

Using Eqs.~(\ref{H_v1v2}), and taking into account the re-scaling of the Hamiltonians $H^{(1)}$ and $H^{(2)}$ by $\hbar \omega_A$, we obtain their common eigenvalues in the harmonic approximation,
\begin{equation}
  \epsilon_{QMT} (m,n) = \hbar \omega_A \left[ \sqrt{\lambda_1} \left( m+\frac{1}{2} \right) + \sqrt{\lambda_2} \left( n+\frac{1}{2} \right) \right] .
  \label{eigenv_QMT}
\end{equation}
When $\lambda_1 \gg \lambda_2$ (see Fig.~\ref{lambda1_lambda2}), mostly the first term will contribute to the ground state energy of the particle, whereas the second term may play a role in the thermal excitation, if the device works at temperatures comparable to $\sqrt{\lambda_2} \hbar \omega_A/k_B$, but much smaller than $\sqrt{\lambda_1} \hbar \omega_A/k_B$.

If the system is prepared in the  ground state in one of the local minima of the potential, then the excitation energy required to get into the running state, over the potential barrier (see Fig.~\ref{U_2d}) is calculated from Eqs.~(\ref{eps_ph_1}) and~(\ref{eigenv_QMT}) to be
\begin{equation}
  \epsilon_{ph} 
  =  \epsilon_{ph}^{(1)} - \epsilon_{QMT} (0, 0) \equiv \Delta U . \label{eps_ph_2}
\end{equation}

For typical values of the parameters $I_c = 1~{\rm \mu A}$ and $C_A = C_J = 1$~pF~\cite{ApplPhysLett.103.142605.2013.Oelsner, PhysRevApplied.7.014012.2017.Oelsner}, choosing $\omega_A / (2\pi) = 30$~GHz, from Eq.~(\ref{eps_ph_2}) we obtain $I/I_c \approx 0.8523$ (so, the bias current is quite close to the critical current).
To be able to detect photons of energy $\hbar \omega_A$, the temperature $T$ of the system and of the environment should be significantly lower than $T_A \equiv \hbar \omega_A / k_B \approx 1.4398$~K.
The angular frequencies of the two independent oscillators of Eq.~(\ref{eigenv_QMT}) are $\omega_1 \equiv \lambda_1 \omega_A$ and $\omega_2 \equiv \lambda_2 \omega_A$.
For the chosen parameters, $\lambda_1 = 2.0226$ and $\lambda_2 = 0.0221$, so, if $T\ll \hbar \omega_A / k_B$, the oscillator 1 should be in the ground state at temperature $T$. Nevertheless, since $\lambda_2 \ll \lambda_1$ the oscillator 2 could be in an excited state if $T$ is comparable or higher than $T_2 = \lambda_2 \hbar \omega_2 / k_B$.
To check the harmonic approximation, we compare $\epsilon_{ph}$ with $\epsilon_{ph}^{(1)}$ and we obtain $\epsilon_{QMT}(0,0)/(\hbar\omega_A) \approx 1.0080$.
This implies $\epsilon_{QMT}(0,0)/\epsilon_{ph}^{(1)} \approx \epsilon_{ph}/ \epsilon_{ph}^{(1)} \approx 0.5$, so the harmonic approximation is reasonably well justified.

The choices of parameters of the JJ, which meet the criterion~(\ref{eps_ph_2}) is broad.
For example, if the critical current is $I_c = 13.7~{\rm \mu A}$~\cite{ApplPhysLett.103.142605.2013.Oelsner, PhysRevApplied.7.014012.2017.Oelsner}, then, from Eq.~(\ref{eps_ph_2}) we obtain $I/I_c = 0.9729$, so, the relative value of the bias current is much closer to one than in the previous case.
In this case, $\lambda_1 \approx 2.3523$ and $\lambda_2 \approx 0.2605$, so, as can be seen in Fig.~\ref{lambda1_lambda2}, the ratio $\lambda_1/\lambda_2$ is not as big as for $I_c = 1~{\rm \mu A}$.
Furthermore, $\epsilon_{QMT}(0,0)/(\hbar\omega_A) \approx 1.1094$ and $\epsilon_{QMT}(0,0)/\epsilon_{ph}^{(1)} \approx 0.5259$, so the harmonic approximation is, again, quite well justified.

\subsection{Dark counts} \label{subsec_dark}

In the absence of the additional energy coming from the photon or from the thermal bath, the easiest escape route for the phase particle from the potential well is to tunnel through the saddle point region.
To estimate the escape rate, we write the Hamiltonian $H^{(2)}$ in the variables $\phi_a$ and $\phi_b$~(\ref{vars_f1_f2_H2}) and emphasize the washboard potential.
In these variables, Eq.~(\ref{H2_ad_v2}) becomes
\begin{eqnarray}
  \hat H_d^{(2)} (\phi_a, \phi_b) &=& - \frac{1}{2} \frac{\partial^2}{\partial\phi_a^2}
  - \frac{1}{2} \frac{\partial^2}{\partial\phi_b^2}
   + \frac{1}{2} \left( 1+\frac{C_A}{C_J} \right) \phi_a^2
  + \frac{I_c}{2e \omega_A} \left[1 - \cos\left( \frac{-\sqrt{C_A/C_J} \phi_a + \phi_b}{\sqrt{Z_0 C_J \omega_A (1 + C_A/C_J)}} \right) \right] \nonumber \\
  && - \frac{I}{2e \omega_A} \frac{-\sqrt{C_A/C_J} \phi_a + \phi_b}{\sqrt{Z_0 C_J \omega_A (1 + C_A/C_J)}}.
   \label{H2_phiab}
\end{eqnarray}
%
%
From Eqs.~(\ref{U2_min_C1}) and (\ref{U2_sad_C1}) we see that the local minima and the saddle points are located along $\phi_b$ axis ($\phi_a = 0$).
Along this axis ($\phi_a = 0$) -- multiplying the Hamiltonian by $\hbar\omega_A$, to go back to $H^{(2)}$ in energy units and defining $\tilde\phi_b \equiv 2e/\sqrt{2C_J \hbar\omega_A (1 + C_A/C_J)}$ -- we obtain
\begin{eqnarray}
  \hat H^{(2)} (0,\phi_b) 
  %
  &=& - \frac{(2e)^2}{2C_J (1 + C_A/C_J)} \frac{\partial^2}{\partial \tilde\phi_b^2}
  - E_c \left[\frac{I}{I_c} \tilde\phi_b + \cos\left( \tilde\phi_b \right) \right]
  + E_c ,
  \label{H2_phia_0}
\end{eqnarray}
which describes a (phase) particle in a washboard potential.
We notice that the ``mass`` of this particle became bigger than the mass of the Josephson particle by $1+C_A/C_J \ (\approx 2$ for our choice of parameters), due to the presence of the antenna.
In such a situation, we shall estimate the order of magnitude of the escape rates as usual, in the model of Caldeira and Leggett~\cite{PhysRevLett.46.211.1981.Caldeira}, considering the tunneling along this axis.
A more detailed analysis of the escape rate in a 2D potential will be done elsewhere.

The escape rate, taking into account the dissipation, may be calculated as~\cite{PhysRevLett.46.211.1981.Caldeira, ApplPhysLett.103.142605.2013.Oelsner}
\begin{equation}
  \Gamma = \frac{\omega_0}{2\pi} \sqrt{\frac{B}{2\pi}} e^{-B} , \label{Gamma_Kuzmin}
\end{equation}
where $B = \Delta U/(\hbar\omega_0)(7.2+8A/Q)$, $Q = \omega_0 RC$ is the quality factor, $A$ is a numerical parameter, $R$ is the shunt resistance of the junction, and $\Delta U = \hbar\omega_A$ is the height of the potential barrier at the saddle point given by Eq.~(\ref{eps_ph_2}); the angular frequency $\omega_0 = \omega_A \sqrt{\lambda_2}$ corresponds to the local minimum of the 1D potential~(\ref{H2_phia_0}).
Using the numerical values from~\cite{ApplPhysLett.103.142605.2013.Oelsner} for an estimation ($R = 0.44~{\rm k\Omega}$ and $A=10$), we obtain
\begin{equation}
  \Gamma_1 \approx 1.8 \times 10^{-30}~{\rm Hz}
\end{equation}
for  the first case ($I_c = 1~{\rm \mu A}$) and
\begin{equation}
  \Gamma_2 \approx 1.7 \times 10^{-2}~{\rm Hz}
\end{equation}
for the second case ($I_c = 13.7~{\rm \mu A}$).
We observe that only in the first case the escape rate is low enough to allow a counting rate of one photon in a few hours.
Therefore, the escape rate imposes an additional constriction for the choice of the JJ parameters, beside Eq.~(\ref{eps_ph_2}).

\section{Conclusions} \label{sec_conclusions}

We studied theoretically the possibility of counting microwave photons with a detector consisting of a Josephson junction (JJ) coupled to an antenna.
In this paper, we focus on derivation of the effective Hamiltonian of the device. Its quantum dynamics including an interplay between activation and tunneling, as well as thermalization after an absorption event, will be considered elsewhere. Such an analysis will allow the estimation the ``dead time" of the detector after a counting of a photon.

We analyzed two configurations.
In the first one, the JJ is connected in series with the antenna and is similar to the photon counter with cold electron bolometer (CEB)~\cite{ApplPhysLett.82.293.2003.Anghel, arXiv181105326.Anghel}, in which the CEB is replaced by the JJ.
In the second configuration, the JJ is capacitively coupled to the antenna and is similar to the one used in~\cite{PhysRevA.89.033853.2014.Andersen}, for example.
For the two configurations, we constructed simple equivalent electric circuits and we wrote the Hamiltonians in the macroscopic quantum tunneling (MQT) formalism.
In this formalism, the dynamics of the JJ is described as a quantum particle moving in a two-dimensional (2D) potential landscape.
After appropriate changes of coordinates, the Schr\"odinger equations describing the two setups were put into equivalent forms, which implies that they have the same quantum mechanical properties.

The potential landscape in which the quantum particle is moving has the shape of a washboard potential in one direction and is parabolic in another direction--these two directions are not perpendicular to each-other (see Eqs.~\ref{QM_U1} and~\ref{QM_U2}).
If the bias current is smaller than the critical current of the junction, the washboard potential forms (as  in the one-dimensional case) an infinite chain of local minima, in which the quantum particle can be placed in an initial metastable state (see Fig.~\ref{U_2d}).
The local minima are separated by potential barriers and the minimum of each potential barrier is a saddle point.

To calculate analytically the metastable states (especially the ``ground state''), in the vicinity of a local minimum we approximate the potential energy by a harmonic potential.
At low temperatures, the particle is assumed to be initially in the ground state of a local minimum.
When the photon is absorbed, the particle is excited.
If the excitation energy is bigger than the difference between the energy of the saddle point and the ground state energy in the closest local minimum, then the system may get into the running state and the photon is detected.

We calculated the parameters of the device which would allow it to detect 1~cm wavelength photons.
If the critical current of the junction is $I_c = 1~{\rm \mu A}$ and $C_A = C_J = 1$~pF, then the bias current $I$ should satisfy the condition $0.8523 I_c \le I < I_c$ to ensure that the photon is able to excite the particle over the potential barrier.
As another example, we took $I_c = 13.7~{\rm \mu A}$ and in this case (keeping $C_A = C_J = 1$~pF) we obtain $0.9729 I_c \le I < I_c$ to be able to detect the 1~cm wavelength photon.

The photon energy $\epsilon_{ph}$ does not have to be larger than the potential energy barrier $\Delta U$ to be detected.
By exciting the JJ to a state of higher energy, even if that energy is lower than the potential barrier, the phase particle may get into the running state by tunneling into the continuum.
The tunneling rate increases rapidly with the energy of the phase particle.

The tunneling process may cause also \textit{dark counts}, when the phase particle tunnels the potential barrier directly from the ground state, that is, in the absence of a photon.
This has been estimated in Section~\ref{subsec_dark} and it turned out that only in the first case ($I_c = 1~{\rm \mu A}$) the dark count rate is low enough to permit the observation of one relevant photon (produced by axion's decay) in a few hours, whereas in the  second case ($I_c = 13.7~{\rm \mu A}$) the dark count rate is of the order of 1/minute, so the detector may not be used for axions detection.
Therefore, even if the parameters of the junction and the bias current may be tuned so that the photon excites the phase particle above the saddle point energy, the dark count rates impose stringent conditions in setting up the working point of the JJ axion (microwave photon) counter.
Nevertheless, in experiments in which the flux of incoming microwave photons is higher (so that the dark count rate does not have to be so low), the JJ single photon counter offers much more flexibility.

\section*{Acknowledgements}

The work was supported by the UEFISCDI (project PN-19060101/2019) and the Romania-JINR collaboration projects, positions 23, 24, 26, Order 397/27.05.2019 (IFIN-HH) and Russian Science Foundation (Project No. 16-19- 10468).

\end{document}